\documentclass[conference,a4paper]{IEEEtran}
\IEEEoverridecommandlockouts

\usepackage[dvipdf]{graphicx}
\usepackage{subfigure}
\usepackage{url}
\usepackage{fancyhdr}
\usepackage{eclbkbox}
\usepackage{indent}
\makeatletter
\def\tbcaption{\def\@captype{table}\caption}
\def\figcaption{\def\@captype{figure}\caption}

\begin{document}
\title{Recommendation System of Grants-in-Aid \\for Researchers by using JSPS Keyword
\thanks{\copyright 2015 IEEE. Personal use of this material is permitted. Permission from IEEE must be obtained for all other uses, in any current or future media, including reprinting/republishing this material for advertising or promotional purposes, creating new collective works, for resale or redistribution to servers or lists, or reuse of any copyrighted component of this work in other works.}
}

\author{\IEEEauthorblockN{Shin Kamada}
\IEEEauthorblockA{Dept. of Intelligent Systems\\
Graduate School of Information Sciences\\
Hiroshima City University\\
3-4-1, Ozuka-Higashi, Asa-Minami-ku\\
Hiroshima, 731-3194, Japan\\
da65002@e.hiroshima-cu.ac.jp}
\and
\IEEEauthorblockN{Takumi Ichimura}
\IEEEauthorblockA{Faculty of Management and Information Systems\\
Prefectural University of Hiroshima\\
\\
1-1-71, Ujina-Higashi, Minami-ku\\
Hiroshima, 734-8559, Japan\\
ichimura@pu-hiroshima.ac.jp}
\and
\IEEEauthorblockN{Takanobu Watanabe}
\IEEEauthorblockA{Community Liaison Center\\
Prefectural University of Hiroshima\\
\\
1-1-71, Ujina-Higashi, Minami-ku\\
Hiroshima, 734-8559, Japan\\
t-watanabe@pu-hiroshima.ac.jp}
}

\maketitle

\pagestyle{fancy}{
\fancyhf{}
\fancyfoot[R]{}}
\renewcommand{\headrulewidth}{0pt}
\renewcommand{\footrulewidth}{0pt}

\begin{abstract}
  An acquisition of a research grant is important for the researchers to conduct a research. The university will build up the organization and reinforce the acquirement of external funds. The researcher becomes aware of grant information and should investigate what kinds of grant it is. Therefore, the staff at the support center for the Industry-Academia collaboration will classify the grant into some categories according to the research fields. However, the task is difficult to realize the matching of the research fields, because the expert knowledge is required to completely classify them. We have developed recommendation system of Grant-in-Aid system for researchers by using JSPS (Japan Society for the Promotion of Science) keywords. The characteristic keywords are extracted from web sites and then the association rules between researchers and grants are determined in the IF-THEN rule format. This paper discusses the experimental results by using the developed system.
\end{abstract}

\begin{IEEEkeywords}
Recommendation system, Grants-in-Aid for researchers, Japan Society for the Promotion of Science (JSPS) Keyword, Association Analysis
\end{IEEEkeywords}

\IEEEpeerreviewmaketitle

\section{Introduction}
\label{sec:Introduction}
Advances in recent information technology enable to collect various database not only numerical values but also comments, numerical evaluation, and binary data such as pictures. The technical methods to discover knowledge in such databases are known to be a field of data mining and developed in various research fields. Association Analysis \cite{Watanabe} is one of famous data mining methods which find interesting knowledge in large database.

An acquisition of a research grant is important for the researchers to conduct a research. The university will build up the organization and reinforce the acquirement of external funds. The researcher becomes aware of grant information and should investigate what kinds of grant it is. Therefore, the staffs at the support center for the Industry-Academia collaboration will classify the grant into some categories according to the research fields, then recommend them each department. However, the task is difficult to realize the matching of the research fields, because the expert knowledge is required to completely classify them. In addition to this, there is a case that some grant information don't have enough amount of information to classify because it is paper based poster or email. In this case, the staffs must refer additional information such as Web site. Moreover, it takes much time for classification that is suitable for an individual because of self-classification. But, researcher fields may be different even if they belong to same department. In this way, it is very important problem that the matching between grants information and researchers in university don't well work.

In this paper, we developed matching system between grants information and researchers by using JSPS (Japan Society for the Promotion of Science) \cite{JSPS-keyword} keyword \cite{Kamada15}. The characteristic keywords with high frequency are extracted from Web site of a grant organization and then JSPS keyword table is used to determine the researcher field of a grant information. Morover, association rules with strong relationships between the characteristic keywords can be extracted by Association Analysis. This paper explains about our developed recommendation system and discusses some experimental results.

The remainder of this paper is organized as follows. In Section \ref{sec:AssociationAnalysis}, the basic concept of Association Analysis and Apriori algorithm which is the well-known algorithm of Association Analysis will be explained briefly. Section \ref{sec:DevelopedSystem} describes our developed recommendation system of Grants-in-Aid by using JSPS keyword. Section \ref{sec:ExperimentalResults} describes some experimental results. In Section \ref{sec:Conclusion}, we give some discussions to conclude this paper.

\section{Association Analysis}
\label{sec:AssociationAnalysis}
This section describes the basic concept of Association Analysis \cite{Watanabe} and Apriori Algorithm \cite{Agrawal93} which is fast algorithm for them.

Association Analysis \cite{Watanabe} is one of famous data mining methods, which is often used to discover hidden relationships in large database. Table \ref{tab:transaction} shows an example of database in Association Analysis. A database consists of some transactions, each transaction has some items (the database of Table \ref{tab:transaction} has 4 transaction and 5 kinds of items). Association Analysis is try to find frequent association rules between items with strong relationships, each rule is represented $(X \Rightarrow Y)$ such as IF-THEN rule (it means that if item $X$ appear in a transaction, then $Y$ will be also appear in same transaction. ). 

In Association Analysis, the strength of relationships of association rule is defined by several way, ``Support'' and ``Confidence'' are well-known one. ``Support'' shows the frequency in database, calculated by Eq. (\ref{eq:support}).

\begin{equation}
\label{eq:support}
{\rm supp}(X \Rightarrow Y)=\frac{\sigma (X \cup Y)}{M},
\end{equation}
where, $X$ and $Y$ are items in database, $M$ is the total number of transactions. $\sigma (X \cup Y)$ is the number of transactions that contain both $X$ and $Y$ (for example, ${\rm supp}(\{ {\rm Item 2}\} \Rightarrow \{ {\rm Item 3}\})$ in Table \ref{tab:transaction} is 3/5).

``Confidence'' measures how often items in $Y$ appear in transactions that contain $X$, calculated by Eq. (\ref{eq:confidence}).

\begin{equation}
\label{eq:confidence}
{\rm conf}(X \Rightarrow Y)=\frac{{\rm supp}(X \Rightarrow Y)}{{\rm supp}(X)},
\end{equation}
where, ${\rm supp}(X)$ is the proportion of transactions that contain $X$ in database (for example, ${\rm conf}(\{ {\rm Item 2}\} \Rightarrow \{ {\rm Item 3}\})$ in Table \ref{tab:transaction} is 3/4). The value of ``Support'' and ``Confidence'' is higher, it means that the association rule is more frequent.

\begin{table}[tbp]
\caption{An example of database}
\label{tab:transaction}
\begin{center}
\scalebox{1.2}[1.2]{
\begin{tabular}{c|l}
\hline \hline
Transaction ID   &  Item　Sets\\ \hline\hline
1                 &  \{Item 1, Item 2\} \\\hline
2                 &  \{Item 2, Item 3, Item 5\} \\\hline
3                 &  \{Item 2, Item 3\} \\\hline
4                 &  \{Item 4\} \\ \hline
5                 &  \{Item 2, Item 3\} \\
\hline \hline
\end{tabular}
} 
\end{center}
\end{table}

However, there is computational complexity in large database because the number of possible configurations of association rules is large according to the database size. The fast algorithm to solve this problem was Apriori Algorithm, introduced by Agrawal et.al \cite{Agrawal93}. Apriori Algorithm has 2 basic principle: one is to define minimum threshold for ``Support'' and ``Confidence'', the other is that if an item set is frequent, then all of its subsets must also be frequent. Fig. \ref{fig:Apriori} shows the procedure of Apriori Algorithm．

\begin{center}
%\begin{indentation}{1.2zw}{1.2zw}
\begin{indentation}{0.0cm}{0.2cm}
\begin{breakbox}
\smallskip
\begin{description}
\item[Step 1) ] Generate candidate item set $C_1$ that contains one item.
\item[Step 2) ] Generate frequent item set $L_1$ with min\_support in $C_1$. min\_support is the minimum threshold for ``Support''.
\item[Step 3) ] Generate candidate item set $C_k$ from $L_{k-1}$. Initial value of $k$ is 2.
\item[Step 4) ] Generate frequent item set $L_k$ with min\_support in $C_k$.
\item[Step 5) ] Add 1 to $k$ and the procedure from Step 3) to Step 4) is executed until candidate item set $C_k$ cannot be generated.
\end{description}
\end{breakbox}
\end{indentation}
\figcaption{The procedure of Apriori Algorithm}
\label{fig:Apriori}
\end{center}

\section{Development of Recommendation System of Grants-in-Aid\\by using JSPS Keyword}
\label{sec:DevelopedSystem}

\subsection{JSPS Keyword \cite{JSPS-keyword}}
In this section, we explain about our developed recommendation system of Grants-in-Aid by using JSPS Keyword. JSPS is Japan Society for the Promotion of Science, the keywords are defined for each researcher fields. In our system, JSPS keywords are used to estimate researcher fields of grants information and researchers.

Table \ref{tab:JSPSkeyword} shows an example of JSPS Keyword. It consists of 4 columns: Area, Discipline, Researcher Field and Keyword. Area, Discipline and Researcher Field means the research category, Keyword is the word related to the researcher fields. The total number of Area, Discipline, Researcher Field and Keyword is 14, 80, 322 and 3674, respectively (about 11 keywords are defined in each researcher field). Our developed system can calculate how appear JSPS keyword with strong relationships in one Web site by using Association Analysis.

\begin{table*}[tbp]
\caption{An example of JSPS Keywords \cite{JSPS-keyword}}
\label{tab:JSPSkeyword}
\begin{center}
\scalebox{1.2}[1.2]{
\begin{tabular}{c|c|c|l}
\hline \hline
Area         &  Discipline                  &  Research Field              & Keyword\\ \hline\hline
Informatics  &  Principles of Informatics   &  Theory of informatics       & Theory of computation, Automata theory, $\cdots$\\
             &                              &  Mathematical informatics    & Optimization theory, Mathematical finance, $\cdots$\\
             &  Human informatics           &  Intelligent informatics     & Machine learning, Knowledge acquisition, $\cdots$\\ \hline
Complex systems &  Human life science       &  Eating habits               & Cooking and processing, Food storage, $\cdots$\\
&                           &  Clothing life/Dwelling life & Dwelling life, Clothing culture, $\cdots$\\ \hline
Engineering & Mechanical engineering      &  Materials/Mechanics of materials               & Continuum mechanics, Structural mechanics, $\cdots$\\
&                           &  Thermal engineering & Thermophysical property, Convection, $\cdots$\\ 
&   Integrated engineering & Aerospace engineering& Aerodynamics, Structure/Material, $\cdots$\\ \hline
$\cdots$ &  $\cdots$  &  $\cdots$ & $\cdots$\\
\hline \hline
\end{tabular}
} 
\end{center}
\end{table*}

\subsection{Developed Recommendation System \cite{Kamada15}}
\label{sec:system}
\begin{figure}[tbp]
\begin{center}
\includegraphics[scale=0.35]{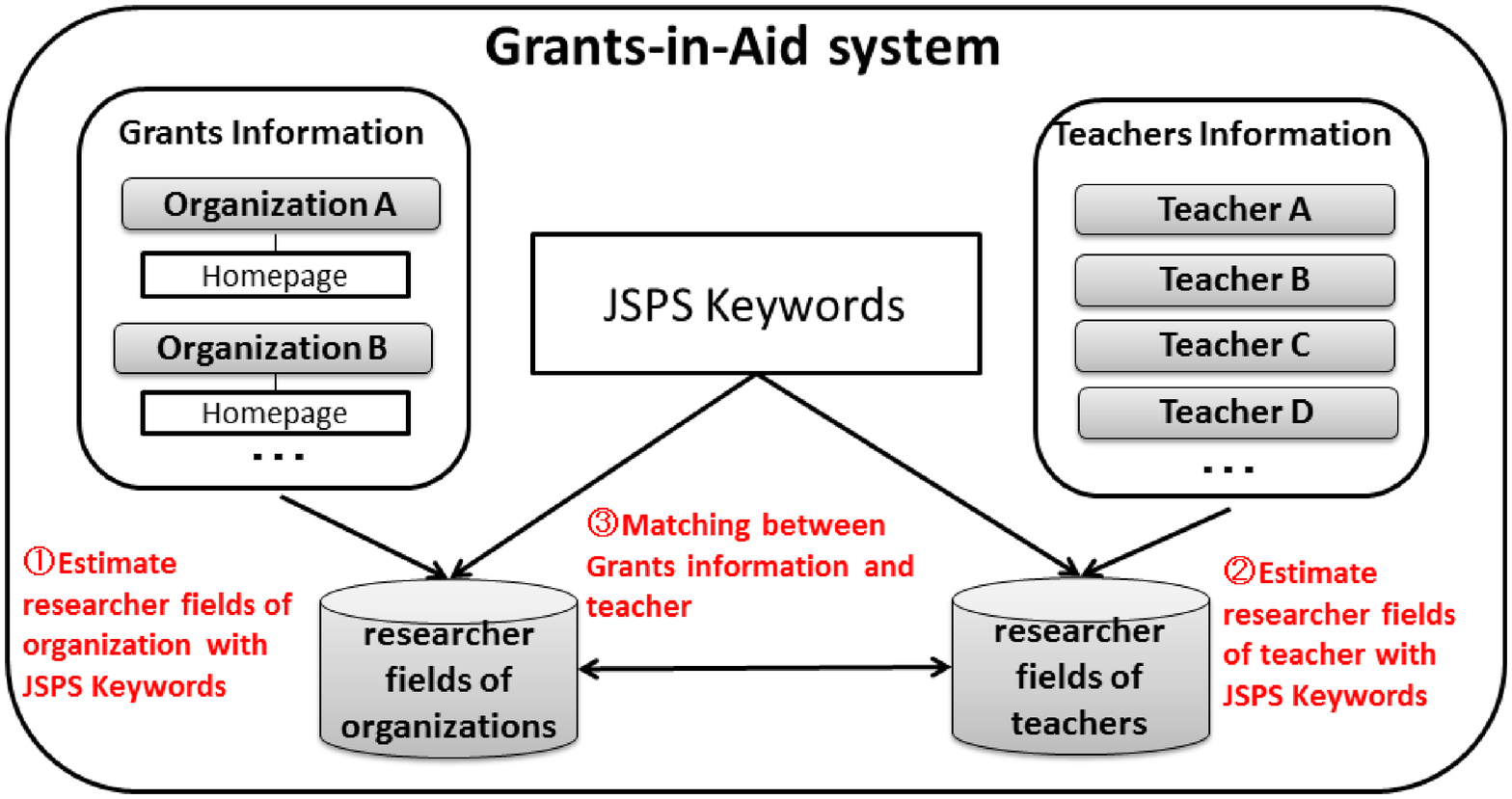}
%\vspace{-3mm}
\caption{The overview of our developed system}
\label{fig:overview}
\end{center}
\end{figure}

Our developed recommendation system of Grants-in-Aid for researchers will be operated as following procedure. Fig. \ref{fig:overview} shows the overview of this system.

\begin{figure}[tbp]
\begin{center}
\includegraphics[scale=0.9]{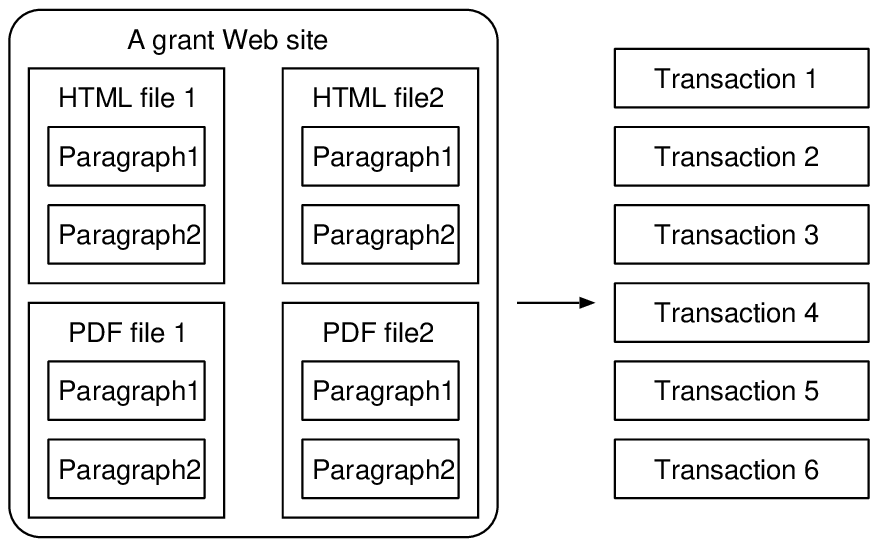}
%\vspace{-3mm}
\caption{How to create transaction}
\label{fig:transaction}
\end{center}
\end{figure}

\begin{enumerate}
\item Download grant information\\
Download PDF and HTML files from Web site in each grant organization by using ``wget'' (Linux shell command).

\item Convert to text files\\
HTML files and PDF files that obtained in Step 1) are converted to text files. Following regular expression is used to remove HTML tags.

\begin{verbatim}
<("[^"]*"|'[^']*'|[^'">])*>
\end{verbatim}

PDF files are converted to text files by using ``pdftotext'' (Linux shell command).

\item Create Transaction\\
Converted text files that obtained in Step 2) are divided into sub-text according to paragraph as shown in Fig. \ref{fig:transaction}. Each sub-text can be seen as a transaction in association analysis.

\item Create Item Sets\\
Some words are extracted from a transaction by using MeCab. MeCab is well-known Japanese Morphological Analysis Engine \cite{MeCab}. Extracted words can be seen as items in the transaction.

\item Extract Association Rules\\
``Support'' and ``Confidence'' are calculated by Eq. (\ref{eq:support}) and Eq. (\ref{eq:confidence}). Then, frequent association rules with ${\rm min\_supp}$ and ${\rm min\_conf}$ are extracted by using Apriori algorithm. ${\rm min\_supp}$ and ${\rm min\_conf}$ are the minimum threshold for ``Support'' and ``Confidence''. In this system, we used the R library called ``arule \cite{arules}'' to calculate Apriori algorithm. Finally, researcher fields are estimated to calculate the most frequent JSPS keywords in acquired association rules.
\end{enumerate}

\subsection{Estimation of researcher fields for teachers}
Our developed system also can estimate researcher fields for researchers in university as well as Section \ref{sec:system}. In Prefectural University of Hiroshima, each researcher data consists of research fields, research title, related keywords and so on. However, they don't have enough amount of information to extract rules by Association Analysis. Therefore, we calculate the most frequent JSPS keywords and its researcher field in researcher data of Prefectural University of Hiroshima.

\section{Experimental Results}
\label{sec:ExperimentalResults}
In this section, we describe some experimental results. 18 grants information and 42 teachers were used in this experiment. The following parameters were used for Apriori Algorithm: ${\rm min\_supp} =$ average of the total support, ${\rm min\_conf} =0.8$.

Table \ref{tab:result-site-category} shows the estimated researcher fields for 18 grans information. Second column of the table shows the example of acquired association rules. Third column shows the estimated researcher field. 

In Artificial Intelligence Research Promotion Foundation \cite{airpf}, the number of acquired transactions was 2071. Table \ref{tab:transaction_airpf} shows an example of transactions, following words were more frequent: ``Artificial'', ``Intelligence'', ``Robot'', ``Information'' and so on. Table \ref{tab:rule_airpf} shows acquired association rules with high frequency in same organization. 248 rules were acquired as frequent rules with ${\rm min\_supp}$ and ${\rm min\_conf}$. As a result, the research field ``Informatics, Human informatics, Intelligent robotics'' was determined because JSPS keywords in this research field consist of ``Intelligence'', ``Robot'' and so on.

Table \ref{tab:result-researcher-category} shows the summary of estimated researcher fields for 43 researchers in Prefectural University of Hiroshima. For each department, the characterestic researcher field were acquired.

Finally, matching result between grants information and researchers was acquired as show in Table \ref{tab:result-matching}. Second column of the table shows the department which matched researchers belong to. Third columns shows the number of researchers matched with each grant information. About 2 researchers were matched for each grans information in ``Discipline''.

\section{Conclusion}
\label{sec:Conclusion}
This paper presented the our developed Grants-in-Aid system by using JSPS Keyword. By using our developed system, researcher fields of grants information and researchers in university are estimated according to JSPS keyword, matching results can be acquired instead of self-classification. As the experimental result, the matching between 18 grants information and 42 teachers was performed, roughly classification was acquired and then it enable to recommend a grant information that is suitable for a each researcher.

In future, we will improve the performance of the developed system and develop the system operating as Web system.

\begin{table*}[tbp]
\caption{Estimated result for Grans information}
\label{tab:result-site-category}
\begin{center}
\scalebox{1}[1]{
\begin{tabular}{l|l|l}
\hline \hline
Grants Information& Acquired Association Rules& Area, Discipline, Researcher Field\\ \hline\hline
The Society of Yeast Sciences\cite{yeast} &\{Microbiology\}$\rightarrow$\{Drug\}&Agricultural sciences, Agricultural chemistry,\\ 
&\{Agricultural\}$\rightarrow$\{Life\}& Applied microbiology \\ \hline
The Japan Bifidus Foundation\cite{bifidus} &\{Breeding\}$\rightarrow$\{Gene\}&Agricultural sciences, Agricultural chemistry,  \\ 
&\{Division\}$\rightarrow$\{Cell\}& Applied biochemistry\\ \hline
Fuji Foundation for Protein Research\cite{fuji} &\{Food\}$\rightarrow$\{Health\}&Agricultural sciences, Agricultural chemistry, \\ 
&\{Health\}$\rightarrow$\{Nutrient\}&  Food science\\ \hline
The Public Foundation of Elizabeth Arnold-Fuji\cite{fujizaidan} &\{Agricultural\}$\rightarrow$\{Food\}& Agricultural sciences, Agricultural chemistry, \\ 
&\{Industry\}$\rightarrow$\{Agriculture\}& Food science\\ \hline
The Food Science Institute Foundation\cite{ryouken} &\{Nutrient\}$\rightarrow$\{Food\}&Agricultural sciences, Agricultural chemistry, \\ 
&\{Development\}$\rightarrow$\{Food\}& Food science\\ \hline
The Tojuro Iijima Foundation for  &\{Industry\}$\rightarrow$\{Food\}&Agricultural sciences, Agricultural chemistry, \\
Food Science and Technology\cite{iijima}&\{Inspection\}$\rightarrow$\{Quality\}& Food science\\ \hline
The Foundation for Dietary Scientific Research\cite{z-ssk} &\{Eating habits\}$\rightarrow$\{Health\}&Agricultural sciences, Agricultural chemistry, \\ 
&\{Analysis\}$\rightarrow$\{Eating habits\}&Food science \\ \hline
Tobacco Academic Studies Center\cite{tasc} &\{Reaction\}$\rightarrow$\{Stress\}&Agricultural, Plant production and environmental \\ 
&\{Peace of mind\}$\rightarrow$\{Environment\}&  agriculture, Science in genetics and breeding\\ \hline
The Fukuhara Memorial Fund for the &\{Literature\}$\rightarrow$\{English\}&Humanities and Social Sciences, Literature,\\
Studies of English and American Literature\cite{daieikyo}&\{English\}$\rightarrow$\{Trust\}&  Literature in English\\ \hline
Artificial Intelligence Research Promotion Foundation\cite{airpf} &\{Artificial\}$\rightarrow$\{Intelligence\}&Informatics, Human informatics, \\ 
&\{Variable\}$\rightarrow$\{Information\}& Intelligent robotics\\ \hline
Foundation for the Fusion of Science and Technology\cite{fost} &\{Software\}$\rightarrow$\{Learning\}&Informatics,\\ 
&\{Advance\}$\rightarrow$\{Simulation\}& Principles of Informatics, Software \\ \hline
Trust Companies Association of Japan\cite{shintaku} &\{Wealth\}$\rightarrow$\{Formation\}&Social Sciences, Law,\\ 
&\{Life\}$\rightarrow$\{Insurance\}&  New fields of law\\ \hline
The Japan Securities Scholarship Foundation\cite{jssf} &\{Law\}$\rightarrow$\{Economy\}&Social Sciences, Law,\\
&\{Welfare\}$\rightarrow$\{Social\}&  Social law\\ \hline
Heiwa Nakajima Foundation\cite{hnf} &\{Earth\}$\rightarrow$\{International\}&Social Sciences, Law, \\ 
&\{Government\}$\rightarrow$\{Law\}& International law\\ \hline
Japan Educational Mutual Aid Association of  &\{Differentiation\}$\rightarrow$\{Education\}&Social Sciences, Education, \\
Welfare Foundation\cite{nikkyoko}&\{Development\}$\rightarrow$\{Education\}&Sociology of education \\ \hline
Mazak Foundation\cite{mazak} &\{Engineering\}$\rightarrow$\{Machine\}&Engineering, Mechanical engineering,\\ 
&\{Industry\}$\rightarrow$\{System\}& Production engineering/ Processing studies \\ \hline
Obayashi Foundation\cite{obayashi} &\{Molecular\}$\rightarrow$\{Chemistry\}&Chemistry, Applied chemistry,\\
&\{Heat\}$\rightarrow$\{Environment\}& Green/ Environmental chemistry\\ 
\hline \hline
\end{tabular}
} 
\end{center}
\end{table*}

\begin{table*}[tbp]
\caption{An example of acquired transactions in Artificial Intelligence Research Promotion Foundation\cite{airpf} }
\label{tab:transaction_airpf}
\begin{center}
\scalebox{1.2}[1.2]{
\begin{tabular}{c|l}
\hline \hline
Transaction ID   &  Item　Sets\\ \hline\hline
1                 &  \{Artificial, Intelligence, Knowledge, Network, Learning, Control, Information \} \\\hline
2                 &  \{Character recognition, Image, Authentification\} \\\hline
3                 &  \{Automaton, Probability, System, Developement\} \\\hline
4                 &  \{Artificial, Intelligence, Relationship, Information, Network\} \\ \hline
5                 &  \{Human, Interface, Contents\} \\ \hline
6                 &  \{Robot, Sensor, Effectiveness\} \\ \hline
7                 &  \{Intelligence, Information, System\} \\ \hline
8                 &  \{Hybrid, Self-support,Movement, Robot, Vision System\} \\\hline
$\cdots$          &  $\cdots$ \\
\hline \hline
\end{tabular}
} 
\end{center}
\end{table*}

\begin{table*}[tbp]
\caption{Acquired frequent association rules in Artificial Intelligence Research Promotion Foundation\cite{airpf}}
\label{tab:rule_airpf}
\begin{center}
\scalebox{1.2}[1.2]{
\begin{tabular}{l|l|l}
\hline \hline
Association Rules                             &  Support  & Confidence\\ \hline\hline
\{Artificial\} $\rightarrow$ \{Intelligence\} &  0.1033   &  0.94 \\
\{Variables\} $\rightarrow$ \{Information\}   &  0.0038   &  0.88 \\
\{Older person\} $\rightarrow$ \{System\}     &  0.0038   &  0.88 \\
\{Hybrid\} $\rightarrow$ \{Automaton\}        &  0.0024   &  1.00 \\
\{Composition\} $\rightarrow$ \{Voice\}       &  0.0024   &  0.83 \\
\{Design\} $\rightarrow$ \{System\}           &  0.0024   &  0.83 \\
\{Venture\} $\rightarrow$ \{Robot\}           &  0.0010   &  1.00 \\
\{Risk\} $\rightarrow$ \{Knowledge\}          &  0.0010   &  1.00 \\
\{Advance\} $\rightarrow$ \{Information\}     &  0.0009   &  1.00 \\
\hline \hline
\end{tabular}
} 
\end{center}
\end{table*}

\begin{table*}[tbp]
\caption{Estimated result for researchers in Prefectural University of Hiroshima}
\label{tab:result-researcher-category}
\begin{center}
%\scalebox{0.8}[0.8]{
\begin{tabular}{c|l|l|l}
\hline \hline
Department  &  Area & Discipline &Researcher Field \\ \hline\hline
Intercultural Studies   &  Humanities  & Literature, Linguistics & Japanese literature, Literature in English\\
                        &  Social sciences  & Psychology, Sociology & Social psychology, Sociology\\\hline
Health Sciences         &  Complex systems  &  Human life science & Eating habits\\ 
                        &  Agricultural sciences & Agricultural chemistry & Food science\\ \hline    
Management          &  Social sciences & Management & Management \\ 
                    &                  & Economics & Money/ Finance\\ \hline  
Management Information System &  Informatics  & Human informatics, Principles of Informatics & Intelligent informatics, Information security \\ 
                              &  Engineering  & Electrical and electronic engineering & Control engineering/ System engineering\\ 
\hline \hline
\end{tabular}
%} 
\end{center}
\end{table*}

\begin{table*}[tbp]
\caption{Matching result between grants information and researchers}
\label{tab:result-matching}
\begin{center}
\scalebox{1}[1]{
\begin{tabular}{l|l|l}
\hline \hline
Grants Information& Matched deparcher & The number of matched researcher\\ \hline\hline
The Society of Yeast Sciences\cite{yeast} & Health Sciences & 1 (matched in Discipline)\\ \hline
The Japan Bifidus Foundation\cite{bifidus} & Health Sciences & 1 (matched in Discipline)\\ \hline
Fuji Foundation for Protein Research\cite{fuji} & Health Sciences & 2 (matched in Discipline)\\
&& 1 (matched in Researcher Field) \\ \hline
The Public Foundation of Elizabeth Arnold-Fuji\cite{fujizaidan} & Health Sciences & 2 (matched in Discipline)\\
&& 1 (matched in Researcher Field)\\ \hline
The Food Science Institute Foundation\cite{ryouken} & Health Sciences & 2 (matched in Discipline)\\
&& 1 (matched in Researcher Field)\\ \hline
The Tojuro Iijima Foundation for  & Health Sciences & 2 (matched in Discipline)\\
Food Science and Technology\cite{iijima}& & 1 (matched in Researcher Field)\\ \hline
The Foundation for Dietary Scientific Research\cite{z-ssk} & Health Sciences & 2 (matched in Discipline)\\
&& 1 (matched in Researcher Field)\\ \hline
Tobacco Academic Studies Center\cite{tasc} &  none & none\\ \hline
The Fukuhara Memorial Fund for the & Intercultural Studies &  8 (matched in Discipline)\\
Studies of English and American Literature\cite{daieikyo}& & 2 (matched in Researcher Field) \\ \hline
Artificial Intelligence Research Promotion Foundation\cite{airpf} & Management Information System & 2 (matched in Discipline)\\ \hline
Foundation for the Fusion of Science and Technology\cite{fost} & Management Information System & 2 (matched in Discipline)\\ \hline
Trust Companies Association of Japan\cite{shintaku} & none & none \\ \hline
The Japan Securities Scholarship Foundation\cite{jssf} & none & none\\ \hline
Heiwa Nakajima Foundation\cite{hnf} & none & none\\ \hline
Japan Educational Mutual Aid Association of  & none & none\\ 
Welfare Foundation\cite{nikkyoko} && \\ \hline
Mazak Foundation\cite{mazak} & Management Information System & 2 (matched in Discipline)\\ \hline
Obayashi Foundation\cite{obayashi} & Health Sciences & 2 (matched in Discipline)\\ 
\hline \hline
\end{tabular}
} 
\end{center}
\end{table*}

\section*{Acknowledgment}
This work was supported by 2014 Priority Research Program assigned by University President of Prefectural University of Hiroshima.

% that's all folks

\begin{thebibliography}{1}
\bibitem{Watanabe}
T.Watanabe, \emph{A Study on Fast Algorithm for Extracting Fuzzy Association Rules}, In Proc. of 26th Fuzzy System Symposium, pp.349-350, 2010.

\bibitem{JSPS-keyword}
Japan Society for the Promotion of Science (JSPS) Keywords, \url{https://www-kaken.jsps.go.jp/kaken1/keywordList.do#top}, [online], 2015.

\bibitem{Kamada15}
S.Kamada, T.Ichimura, T.Watanabe, \emph{Development of Grants-in-Aid system by using JSPS Keyword}, 2015 IEEE SMC Hiroshima Chapter Young Researchers WorkShop，pp.121-124, 2015.

\bibitem{Agrawal93}
R.Agrawal, T.Imielinski and A.Swami, \emph{Mining association rules between sets of items in large databases}, In Proc.of the 1993 ACM SIGMOD International Conference on Management of Data, pp.207-216, 1993.

\bibitem{MeCab}
MeCab, \url{http://mecab.googlecode.com/svn/trunk/mecab/doc/index.html}, [online], 2006.

\bibitem{arules}
H.Michael, G.Bettina, and H.Kurt, \emph{A Computational Environment for Mining Association Rules and Frequent Item Sets}, Journal of Statistical Software, Vol.14, Issue 15, 2005.

\bibitem{yeast}
The Society of Yeast Sciences, \url{http://www.yeast.umin.jp}, [online], 1959.

\bibitem{bifidus}
The Japan Bifidus Foundation, \url{http://bifidus-fund.jp/index.shtml}, [online], 2008.

\bibitem{fuji}
Fuji Foundation for Protein Research, \url{http://www.fujifoundation.or.jp}, [online], 1979.

\bibitem{fujizaidan}
The Public Foundation of Elizabeth Arnold-Fuji, \url{http://www.fujizaidan.or.jp}, [online], 1967.

\bibitem{ryouken}
The Food Science Institute Foundation, \url{http://www.ryouken.or.jp}, [online], 2005.

\bibitem{iijima}
The Tojuro Iijima Foundation for Food Science and Technology, \url{http://www.iijima-kinenzaidan.or.jp}, [online], 2015.

\bibitem{z-ssk}
The Foundation for Dietary Scientific Research, \url{http://www.z-ssk.org/}, [online], 2006.

\bibitem{tasc}
Tobacco Academic Studies Center, \url{http://www.tasc.or.jp}, [online], 2012.

\bibitem{daieikyo}
The Fukuhara Memorial Fund for the Studies of English and American Literature, \url{http://daieikyo.jp/aetp/modules/bulletin/index.php?page=article\&storyid=6}, [online], 2009.

\bibitem{airpf}
Artificial Intelligence Research Promotion Foundation, \url{http://www.airpf.or.jp/}, [online], 2012.

\bibitem{fost}
Foundation for the Fusion of Science and Technology, \url{http://www.fost.or.jp/}, [online], 1994.

\bibitem{shintaku}
Trust Companies Association of Japan, \url{http://www.shintaku-kyokai.or.jp/}, [online], 1997.

\bibitem{jssf}
The Japan Securities Scholarship Foundation, \url{http://www.jssf.or.jp/}, [online], 1973.

\bibitem{hnf}
Heiwa Nakajima Foundation, \url{http://hnf.jp/}, [online], 1959.

\bibitem{nikkyoko}
Japan Educational Mutual Aid Association of Welfare Foundation, \url{http://www.nikkyoko.or.jp/}, [online], 1959.

\bibitem{mazak}
Mazak Foundation, \url{http://www.mazak-f.or.jp}, [online], 2003.

\bibitem{obayashi}
Obayashi Foundation, \url{http://www.obayashifoundation.org}, [online], 2001.
  
\end{thebibliography}
\end{document}